\documentclass[runningheads]{llncs}
\usepackage{graphicx}
\usepackage{textcomp}
\usepackage{amsmath,amssymb,amsfonts}
\usepackage{xcolor,soul}
\sethlcolor{yellow}
\begin{document}
\title{Between-Tactor Display \\Using Dynamic Tactile Stimuli\thanks{This research was funded by a Small Business Innovation and Research (SBIR) grant W81XWH21C0051 through the Defense Health Agency and technical point of contact, Dr. Christopher Brill, through the Air Force Research Laboratory. The views expressed in this paper are those of the authors and do not reflect the official views or policy of the Department of Defense or its Components. Mention of any specific commercial products, process, or service does not constitute or imply its endorsement, recommendation, or favoring by the United States Government, Department of Defense, or Department of Air Force.}}
%
%

\author{Ryo Eguchi\inst{1}\orcidID{0000-0003-2420-5847}, David Vacek\inst{1}, Cole Godzinski\inst{2},\\ Silvia Curry\inst{2}, Max Evans\inst{2}, and Allison M.\ Okamura\inst{1}\orcidID{0000-0002-6912-1666}}
%
\authorrunning{R.\ Eguchi et al.}
%
\institute{
Stanford University, Stanford, CA 94305, USA
\\ \email{\{eguchir,dvacek,aokamura\}@stanford.edu}
\and Triton Systems, Inc., Chelmsford, MA 01824, USA
\\ \email{\{cgodzinski,scurry,mevans\}@tritonsys.com}
}
\maketitle 
\begin{abstract}
Display of illusory vibration locations between physical vibrotactile motors (tactors) placed on the skin has the potential to reduce the number of tactors in distributed tactile displays.
This paper presents a between-tactor display method that uses dynamic tactile stimuli to generate illusory vibration locations.
A belt with only 6 vibration motors displays 24 targets consisting of on-tactor and between-tactor locations. On-tactor locations are represented by simply vibrating the relevant single tactor. Between-tactor locations are displayed by adjusting the relative vibration amplitudes of two adjacent motors, with either 
(1) constant vibration amplitudes or (2) perturbed vibration amplitudes (creating local illusory motion). 
User testing showed that perturbations improve recognition accuracy for in-between tactor localization. 

\end{abstract}

\section{Introduction}
Torso-worn vibrotactile displays providing spatial cues can be used for directional navigation. In these displays, a vibrotactile motor (tactor) vibrates at a point on the human transverse plane to indicate a subjective direction defined by a vector between the torso center and the tactor location. The torso is an attractive location because it has a large skin area and wearable devices can be easily attached, typically in the form of a belt. Applying tactile feedback on the torso also leaves more functional body parts (e.g., hands, fingers, and feet) free for interaction with the environment. Previous studies have examined localization using a single vibrotactile stimulus (e.g., \cite{cholewiak2004vibrotactile,jones2008tactile}) or recognition of a direction using two successive stimuli (e.g., \cite{van2005presenting,van2005vibrotactile}). Representation of locations or movements between physical tactors would increase the resolution of such a display without increasing the number of tactors \cite{israr2011tactile}. In this paper, we propose a novel between-tactor display using dynamic tactile stimuli to achieve this representation.

\section{Methods}
\textbf{Vibrotactile Patterns: }
We used a vibrotactile belt consisting of an elastic strap and six eccentric rotating mass motors (VZ7AL2B1692082, Vybronics), which we refer to in this paper as tactors. The tactors are evenly spaced at 12 cm and attached to the belt using hook-and-loop fasteners. A microcontroller (Nano 33 BLE, Arduino) and motor driver sends commands to the tactors, and the microcontroller communicates with a laptop PC via a USB cable. A customized MATLAB (2021b, MathWorks) graphical user interface (GUI) commands the tactor voltage (which corresponds to vibration amplitude), displays candidate targets, and records a user's response via joystick movement at 100 Hz.

\begin{figure}[t]
\includegraphics[width=\textwidth]{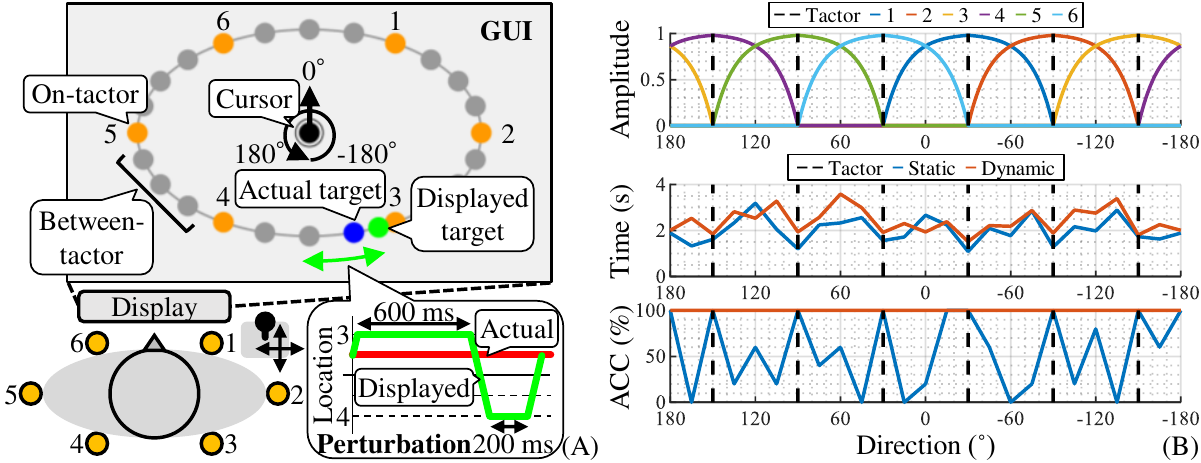}
\vspace{-0.5cm}
\caption{(A) Experimental setup. Only the orange and gray targets are visible to the user. (B) Amplitude of each tactor and demonstration results (time from stimulus onset to target acquisition and recognition accuracy).} \label{fig1}
\vspace{-0.5cm}
\end{figure} 

As shown in Fig.~\ref{fig1}(A), 24 directions are presented using the 6 tactors.
An on-tactor location is represented by vibrating a single tactor.
A between-tactor target location is represented by adjusting relative vibration amplitudes of two adjacent tactors according to their distances to the target. Target locations are defined as an angle indicating the direction of the target from the center of the torso.
The amplitude $y~(\in[0,1])$ with respect to the target angle $x$ is: 
\begin{equation}
    y(x) = 
    \begin{cases}
        \max \left( {1 - \exp \left( { - \frac{{x - d + 60^\circ}}{T}} \right),0} \right) & x \leq d\\
        \max \left( {1 - \exp \left( { - \frac{{d - x + 60^\circ}}{T}} \right),0} \right) & x > d
    \end{cases}
\end{equation}
where $d$ is the angle describing the location of a tactor and $T$ is a time constant and set to 15.
The exponential function maintains sufficient vibration amplitudes of adjacent tactors such that the midpoint between two tactors can be perceived.

During testing, we found that identification of between-tactor locations was poor with constant vibration amplitudes.
To address this, we perturbed the vibration amplitudes for between-tactor locations, creating illusory motion.
The perturbation is a trapezoidal function with a period of 1 s. 
The between-tactor target takes 100 ms to move between the two adjacent tactors and stays on each tactor for a different duration calculated as the inverse of the ratio of distances from that tactor to the between-tactor target
(e.g., 600 ms on the nearest tactor and 200 ms on the other for describing the quantile point).

\textbf{Demonstration:}
We demonstrated the method with a single healthy user. 
The user wore the belt so that two tactors on the front were equally distant from the navel and then learned to identify both on-tactor and between-tactor locations via simultaneous display of a tactile stimulus from the belt and visual indication of its respective location from the GUI.
Twenty-four directions were presented five times each using either static (not perturbed) or dynamic (perturbed) stimuli for the between-tactor locations. On-tactor stimuli were always static.
The participant was asked to use the tactile stimuli to identify target locations, and indicate the perceived location by moving a cursor from the center of the ellipse to targets on the GUI using a joystick (Hotas Warthog Flight Stick, Thrustmaster).

\section{Discussion}
The relative amplitudes of tactors, recognition accuracy, and time from stimulus onset to target acquisition (response time) for each direction are shown in Fig.~\ref{fig1}(B).
Although perturbations slightly prolonged the response time, they significantly enhanced the recognition accuracy for between-tactor locations.
This indicates that the user can quickly and easily identify directional cues from spatiotemporal information, consisting of the illusory motion and differences in tactor vibration duration.
Thus, the proposed between-tactor display can present directional cues with higher resolution than the number of physical tactors. In future work, we will perform a full user study measuring accuracy and response time for various between-tactor resolutions and measure effects of cognitive load.

\bibliographystyle{splncs04}
\bibliography{mybibliography}

\end{document}